\documentclass[traditabstract]{aa}  

\usepackage{natbib}
\bibpunct{(}{)}{;}{a}{}{,} 
\usepackage{graphicx}
\usepackage{txfonts}
\begin{document}

   \title{Contrast sensitivities in the \textit{Gaia} Data Release 2}

   \author{Alexis Brandeker\inst{\ref{inst1}}
          \and
          Gianni Cataldi\inst{\ref{inst2}}\fnmsep\thanks{International Research Fellow of Japan Society for the Promotion of Science (Postdoctoral Fellowships for Research in Japan (Standard)).}
          }
   	\institute{
		Department\ of\ Astronomy,\ Stockholm\ University,\ AlbaNova\ University\ Center,\ 106\,91\ Stockholm,\ Sweden\\
		\email{alexis@astro.su.se}\label{inst1}
		\and
		Subaru\ Telescope,\ National\ Astronomical\ Observatory\ of\ Japan,\ 650\ North\ Aohoku\ Place,\ Hilo,\ HI\,96720,\ USA\label{inst2}
             }


 
  \abstract
   {The source detection sensitivity of \textit{Gaia} is reduced near sources. To characterise this contrast sensitivity is important for understanding
   the completeness of the \textit{Gaia} data products, in particular when evaluating source confusion in less well resolved surveys, such as
   in photometric monitoring for transits. Here, we statistically evaluate the catalog source density to determine the \textit{Gaia} Data Release 2
   source detection sensitivity as a function of angular separation and brightness ratio from a bright source. The contrast sensitivity from $\sim$0.4\arcsec\ out to
   12\arcsec\ ranges in $\Delta G$ = 0--14\,mag. We find the derived contrast sensitivity to be robust with
   respect to target brightness, colour, source density, and \textit{Gaia} scan coverage.}

   \maketitle
%


\section{Introduction}

The final data products from the \textit{Gaia} astrometric mission \citep{pru16} will likely be released well into mid-2020s, but in the mean
time intermediate data releases provide unprecedented high quality in many areas, including astrometry, photometry,  
and completeness \citep{bro18,lin18,eva18,are18}. In addition to serve as a rich source for science exploitation, the \textit{Gaia} data form the basis for source catalogs of
current and future space missions, such as the photometric monitoring missions \textit{TESS} \citep{ric15}, \textit{CHEOPS} \citep{ces17}, and \textit{PLATO} \citep{rau14}. One
key aspect for evaluating the suitability of targets for planning purposes, or the significance of detected variability, is the potential contamination 
of the target by background sources inside an angular 
resolution element (in the range 10--20\arcsec\ for the mentioned photometric missions). Having an estimate 
for the completeness limits of the \textit{Gaia} data near the targets then becomes vitally important for
understanding how bright background stars can be while missing from the catalog.
The contrast sensitivity (CS) is also important when planning e.g.\ high spatial resolution surveys for multiplicity, to be able to predict expected 
companion yields in light of the \textit{Gaia} data.

The pre-launch expected CS of \textit{Gaia} has been estimated for the inner arcsecond \citep{deb15}. Post launch, 
estimates of the completeness of the \textit{Gaia} Data Release 2 (DR2) have been empirically made by comparing \textit{Gaia} detections of multiple systems
to known systems from the Washington Visual Double Star Catalogue \citep[WDS;][]{mas01}. This estimate shows the fraction of WDS companions
recovered in DR2 as a function of separation \citep[Fig.~8 of][]{are18}, but not as a function of brightness contrast. 
Recently, \citet{zie18} made a comprehensive study of the DR2 recoverability of companions detected by a large adaptive optics
survey, and derived the CS as a function of separation and brightness contrast for the inner 3.5\arcsec.
Here, we devise a complementary method to empirically determine the CS out to 18\arcsec\ from source density statistics. Implementing an efficient way of organising the DR2 data, we use the full catalogue to derive the CS. We also divide the sample into subsamples to study how the CS depends on brightness, colour, source density, and \textit{Gaia} scan coverage.


\section{Methods\label{s:methods}}
\subsection{Empirical contrast sensitivity}
We empirically determine the CSs of the DR2 sample by counting the number of sources as a function
of angular distance from a bright source (hereafter called the ``target''), and compare
that with the estimated sky density of those sources at that location in the
sky. Targets are selected to be all sources within the magnitude range $G$ = 6--12\,mag with parallax and proper motion detected with at least 3$\sigma$.
This magnitude range is chosen to include sufficiently many sources for statistics, while at the same time provide a large dynamic range.
For each target a field of 18\arcsec\ radius is checked for background sources. The angular distance $s$ and magnitude difference $\Delta G$ then determine the bin location $(i,j)$ in a two-dimensional
angular separation $i$ vs.\ brightness contrast $j$ grid so that each background source for all targets adds $\Omega_i^{-1}$ to the bin $b_{i,j}$, where $\Omega_i$ is the 
solid angle of ring $i$. Each bin is then normalised by $m_j = \max_i (b_{i,j})$, the maximum accumulated background source density for each $j$. 
Then, $f_{i,j} = b_{i,j} / m_j$ is the fraction of sources detected as a function of angular distance and brightness ratio. 
The source density for a given $\Delta G$ increases sharply at a certain separation and becomes constant at larger separations. 
Due to contamination from physical companions, sources close to the target are in excess compared to the background density. To reduce the 
number of physical companions,  we reject sources with a parallax and proper motion within $3\sigma$ from the target. 
This also removes duplicated sources (described in Sect.~10.2.2 of the DR2 online documentation, release 1.1).
Even so, there remains a significant excess of sources near the target. 
This could be due to spurious detections, which are well known to be present around bright stars \citep[Sect.~6.3 of][]{fab16},
or underestimated errors for sources near bright targets, so that actual companions show spurious parallaxes or proper motions sufficiently different from the target to not be removed. 
By 
using the maximum source density (including remaining physical companions), we effectively assume that the source density is constant within the 
separation of maximum background density, and that the reason for the lower observed density is the reduced sensitivity at close separation.
Since the CS is a strong function of angular separation, the determined limits are robust with respect to variations in true source density. Still, this results in an uncertainty for the CS of the inner 1.5\arcsec\ (briefly addressed in Sect.~\ref{s:discuss}).

In order to efficiently retrieve background sources around targets, we implemented a hierarchical data structure to store the positions and
DR2 identification numbers for all sources. To avoid singularities at the poles, the right ascension and declination of each source define a three-dimensional
directional vector of unit length. An octree of hierarchical cubes is then constructed such that the outermost cube circumscribes the unit sphere, and the direction vector coordinate position
of any source is located in a unique sub-cube. The number of levels of the octree depends on the stellar density but is defined so that no cube contains 
more than a given maximum number $M$ of sources. In our python implementation we find $M = 2 \times 10^5$ to give good performance, resulting in 
at least 8500 cubes at the lowest level for the 
$1.7\times10^9$
sources of the DR2. The octree of cubes makes it efficient to retrieve sources within an angle of any
direction, as only subcubes intersecting the viewing cone need to be considered and directional comparisons can be made
on the cube rather than source level.
By storing the data of sources near each other in the sky
also close to each other on the storage medium, bottle-necking disk access and cache misses are reduced. This software was originally
produced in support of observation preparation and data reduction for the \textit{CHEOPS} mission, but is adapted here for the purpose of characterising the CSs of DR2. 

\subsection{Contrast sensitivity variability\label{s:tests}}
To enable the study of how robust the derived CS is, we divide the DR2 targets into several subsamples and derive
CSs for each subsample separately. For variation with target brightness we divide the sample into targets with 
$G$ = 6--9\,mag and $G$ = 9--12\,mag. The photo-electrons generated by a source is not dependent on the colour of the source, so the
detection threshold in $G$ should in principle be insensitive to the colour of the source. Since the point-spread function (PSF) of the
optical system is colour dependent, however, there may be a colour dependence for the detection threshold, in particular
for the CS where the PSF of the target contributes to the background brightness. We explore the colour dependence by limiting the sample to sources with derived effective temperatures, and dividing it into ``red'' and ``blue'' sources, defined to have effective temperatures cooler and hotter than $T_{\mathrm{break}} = 5000$\,K. 
$T_{\mathrm{break}}$ is chosen to produce approximately equal numbers in each subsample. We then compare the CS for the two cases were the target is either red or blue.

\textit{Gaia} does not scan all parts of the sky equally often, and the number of scans is strongly related to the ecliptic where
regions near the ecliptic pole are covered more than those close to the equator \citep[e.g.\ Fig.~B3\,\&\,B4 of][]{lin18}. In principle,
more scans of a target could improve the CS. To test this, we divide the sample into sources inside and outside 30\degr\ of the ecliptic, corresponding to high and low scan coverage regions.

Since the CS is estimated by normalising by the background source density, 
there should in principle be no dependence on the source density.
We test this by dividing the sample into sources inside and outside 15\degr\ of the galactic plane,
representing high and low source density regions.



\section{Results\label{s:results}}
\begin{figure}
  \centering
   \includegraphics[width=\hsize]{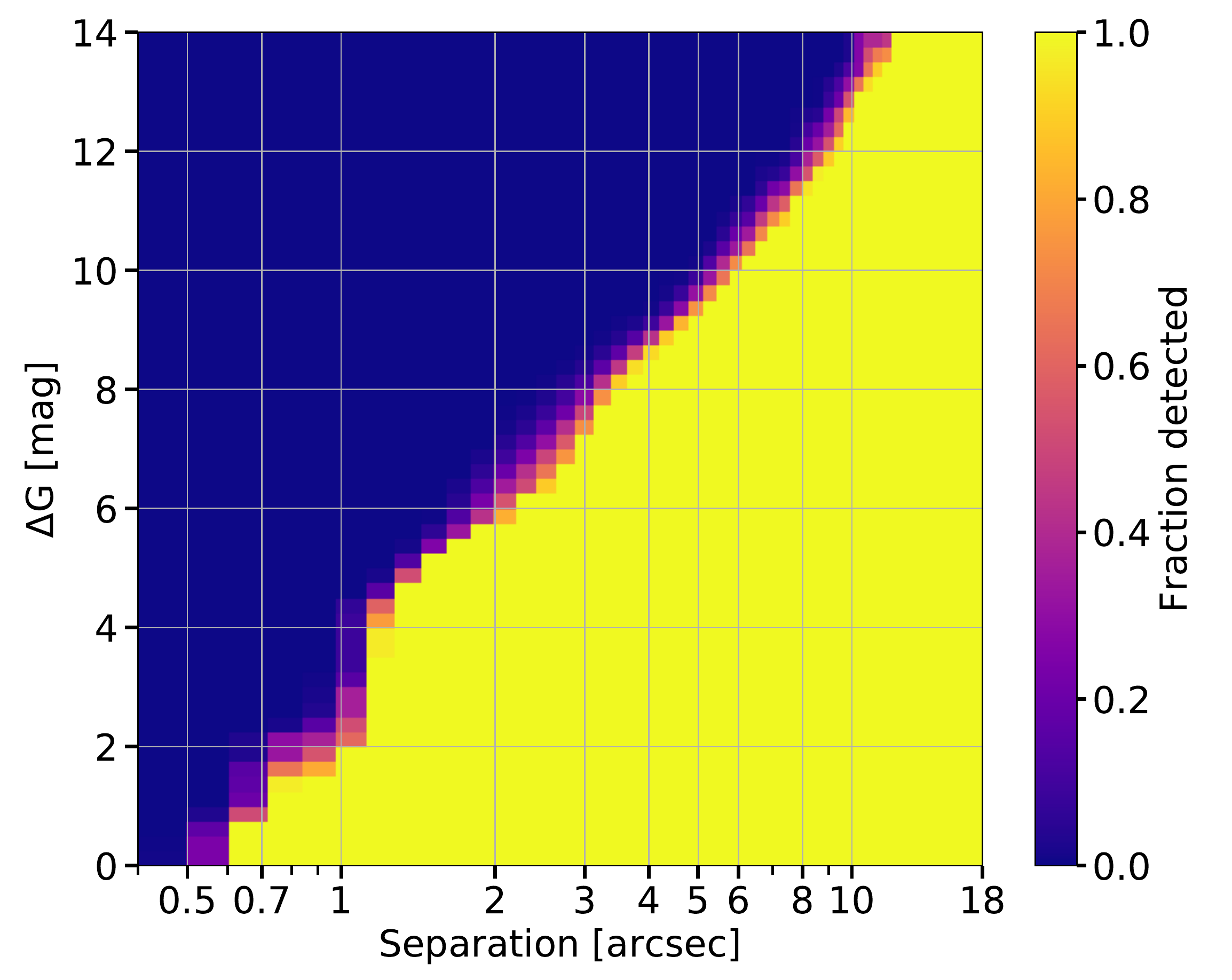}
   \caption{Contrast sensitivity determined as a fraction of detected
   sources as a function of angular distance and magnitude difference
   to a target source. }
      \label{f:contrast}
 \end{figure}
\begin{table}
\caption{Separation as a function of $\Delta G$ and detection fraction}
\label{t:cs}      
\centering
\begin{tabular}{r c c c c c}
\hline\hline
$\Delta G$ & 1\,\%  & 10\,\%  & 50\,\% & 90\,\% & 99\,\%  \\    
(mag) & \multicolumn{5}{c}{(arcsec)}  \\    
\hline                        
 0.5 & 0.46 & 0.50 & 0.59 & 0.65 & 0.66 \\
 1.0 & 0.52 & 0.59 & 0.68 & 0.76 & 0.78 \\
 1.5 & 0.56 & 0.62 & 0.73 & 0.88 & 0.96 \\
 2.0 & 0.59 & 0.69 & 0.94 & 1.1 & 1.1 \\
 2.5 & 0.77 & 0.90 & 1.1 & 1.2 & 1.2 \\
 3.0 & 0.88 & 0.97 & 1.1 & 1.2 & 1.2 \\
 3.5 & 0.93 & 1.1 & 1.1 & 1.2 & 1.3 \\
 4.0 & 0.93 & 1.1 & 1.1 & 1.2 & 1.3 \\
 4.5 & 1.0 & 1.1 & 1.2 & 1.3 & 1.4 \\
 5.0 & 1.2 & 1.3 & 1.4 & 1.5 & 1.5 \\
 6.0 & 1.6 & 1.7 & 2.0 & 2.2 & 2.3 \\
 7.0 & 1.9 & 2.2 & 2.6 & 2.9 & 3.0 \\
 8.0 & 2.4 & 2.8 & 3.2 & 3.5 & 3.6 \\
 9.0 & 3.4 & 3.9 & 4.3 & 4.6 & 4.8 \\
10.0 & 4.8 & 5.1 & 5.6 & 6.0 & 6.1 \\
11.0 & 5.7 & 6.2 & 7.0 & 7.6 & 7.8 \\
12.0 & 7.3 & 7.8 & 8.7 & 9.3 & 9.6 \\
13.0 & 8.6 & 9.2 & 10. & 11. & 11. \\
14.0 & 9.6 & 10. & 12. & 12. & 12. \\
\hline
\end{tabular}
\end{table}
\begin{figure*}
  \centering
   \includegraphics[width=1.0\hsize]{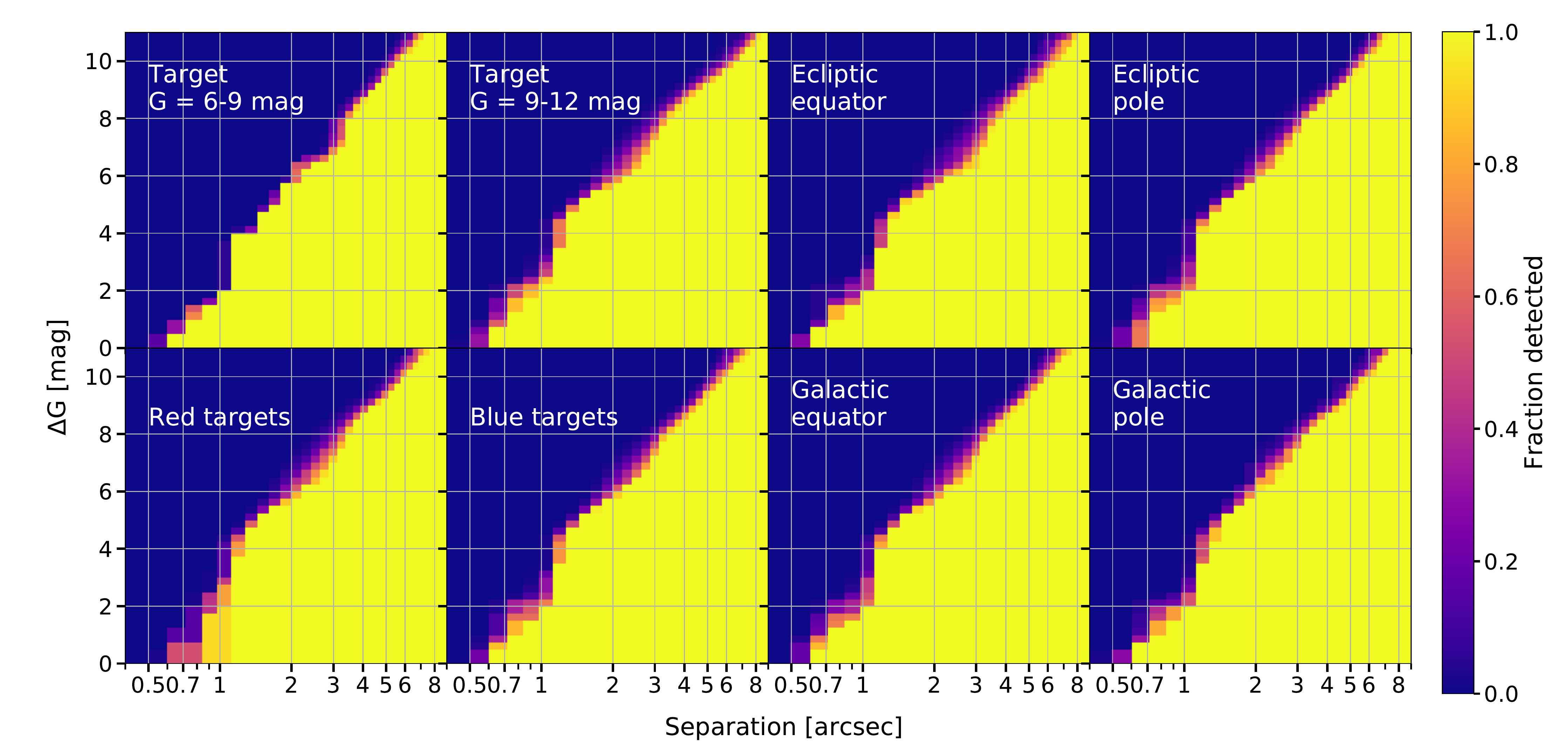}
   \caption{Contrast sensitivities as in Fig.~\ref{f:contrast}, but for a range of subsamples as described in 
   Sect.~\ref{s:tests}. The target brightness range is in general $G$ = 6--12, except for the two samples divided into $G$ = 6--9
   and $G$ = 9--12, as noted in the panels. ``Red'' and ``blue'' targets are defined to be cooler or hotter than
   5000\,K, respectively.
   The ecliptic equator sample is targets within 30\degr\ from the ecliptic plane, while the polar sample is the
   complement. The corresponding dividing angle for the Galactic plane samples is 15\degr.}
      \label{f:test}
 \end{figure*}
The derived CS is shown in Fig.~\ref{f:contrast} as the estimated fraction of detected background sources as a function of 
angular separation (0.4--18\arcsec) and flux ratio ($\Delta G$ = 0--14\,mag). Fig.~\ref{f:contrast} adds up the count of background sources
for targets in the magnitude range 6--12\,mag. Table~\ref{t:cs} tabulates separations from the target, where 1\,\%, 10\,\%,
50\,\%, 90\,\%, and 99\,\% of nearby sources of specified contrast $\Delta G$ to the target, are detected.

The results from the CS variability study is presented in Fig.~\ref{f:test}. In general, the CSs are consistent within $\sim$0.5\,mag. The CSs are smooth with a discontinuity just at 1.1\arcsec. We do not know the origin of this discontinuity, but one plausible 
explanation is that it stems from the DR2 special processing of sources within 2\arcsec. Bright sources are assigned a scanning window of 18 pixels, corresponding to 1.06\arcsec\ (Table~1.1 of the DR2 online documentation, release 1.1); sources closer to each other than this may result in one
source having its scan window truncated, eliminating the source from further processing and thus systematically lowering the detection yield 
of such sources.

The small deviation at the faint end of the $G$ = 9--12\,mag sample is due to the detection limit of 
DR2 (at $G \sim 20$\,mag). The CS of red targets also seem marginally reduced within
1\arcsec, perhaps due to a broader PSF for red targets.


\section{Discussion and conclusions\label{s:discuss}}
We find a significantly lower CS for the inner arcsecond in DR2 compared to the pre-launch
expectations for the full mission \citep[e.g., at 1\arcsec\ we find $\Delta G \sim 3$\,mag compared to their $\Delta G = 7$\,mag;][]{deb15}, 
but are consistent with \citet{zie18} for the inner 3\arcsec\ and
$\Delta G = 7$ they cover. 
This is encouraging, since their method of comparing DR2 to known close-separation sources is not vulnerable to companion contamination. 
The discontinuity we observe at 1.1\arcsec\ is less pronounced in their sample, but it is difficult
to be conclusive at this level of detail due to their limited statistical sample restricting the resolution.

The overabundance of background sources near targets, even after filtering out probable companions by
parallax and proper motion, points to a systematic underestimate of errors at close separations. Since the
overabundance of close sources must be due to either spurious detections or bound companions, the data could
serve as a basis for a statistical study of multiplicity. Because of the special techniques employed for 
registering close sources, a good understanding of potential systematic biases is essential for an accurate
interpretation. To find out if the DR2 errors are biased at separations smaller than a few arcsec, comparing to
a catalogue of known binary systems \citep[like in][]{are18} would likely be a good strategy.

Given current limitations of DR2 in the detection methods applied to sources at small separations \citep{bro18},
future data releases will no doubt improve contrast sensitivities.

\begin{acknowledgements}
We thank the anonymous referee for timely and constructive reports.
AB was supported by the \textit{Swedish National Space Agency}
through contract 75/13. 
\end{acknowledgements}

\bibliographystyle{aa} 
\bibliography{Brandeker_GaiaContrast}

\end{document}